\documentclass{article}
\usepackage[utf8]{inputenc}
\usepackage{authblk}
\usepackage{setspace}
\usepackage[margin=1.25in]{geometry}
\usepackage{graphicx}
\graphicspath{ {./figures/} }
\usepackage{float}
\usepackage{subcaption}
\usepackage{amsmath}
\usepackage{cite}
\usepackage{hyperref}

\title{Electrical Conductivity of Nanoscale Hydrophobic Porous Media: Surface Charge Density and Heterogeneous Pore Structure}

\author[1,2*]{Zhongliang Chen}
\author[1]{Xiaohu Dong}
\author[1,3]{Zhangxin Chen}
\author[2*]{Marc-Olivier Coppens}

\affil[1]{State Key Laboratory of Petroleum Resources and Prospecting, China University of Petroleum (Beijing), Beijing 102249, China.}
\affil[2]{EPSRC “Frontier Engineering” Centre for Nature Inspired Engineering, Department of Chemical Engineering, University College London, London, UK.}
\affil[3]{Department of Chemical and Petroleum Engineering, University of Calgary, Alberta T2N 1N4, Canada.}
\affil[*]{Corresponding author. Email: zhongliang.chen@ucl.ac.uk; m.coppens@ucl.ac.uk}

\date{}

\begin{document}
	
	\maketitle
	
	\begin{abstract}
		Electrical conductivity is an inherent property of a hydrophobic porous media (HPM) and has critical applications. This research aims to provide a solution for predicting the electrical conductivity of nanoscale HPM with heterogeneous pore structure. Molecular dynamics (MD) simulations are compared with the modified Poisson-Boltzmann (MPB) model for understanding ionic charge density distributions in nanopores. The effective medium approximation (EMA) participates in calculating the effective conductance and conductivity of the nanoscale HPM. The results show that the surface charge density affects the ionic density profiles in the hydrophobic nanopores. As the pore size increases, the conductance increases. As the molarity of the aqueous electrolyte solution (AES) decreases, the conductance decreases. A phenomenon related to the conductance saturation occurred when the molarity of AES is very low. The effective conductance of an HPM increase as the coordination number increases. Finally, based on the calculated effective conductance and the heterogeneous pore structure parameters, the electrical conductivity of a nanoscale HPM is calculated.

	\end{abstract}
	
	
	\section{Introduction}
	Unlike conductance, electrical conductivity as an intrinsic property of porous media has essential applications. Examples include measuring the charge transport capacity of electrically conductive porous metal-organic frameworks (MOFs)\cite{Skorupskii2020}, characterizing the geoelectrical properties without destructing the porous rocks and soils\cite{DuyThanh2019}, and participating in the evaluation of the electrochemical reaction rate in porous metals\cite{Ke2007}.
	
	Hydrophobic porous media (HPM) have drawn many research attention due to its excellent liquid transportability\cite{Ye2017,sunden2005transport} and electrical conductivity\cite{Jiang2017}, such as the fabrication of hydrophobic and porous MXene (2D transition metal carbide) foam\cite{Liu2017}. 
	
	Compared with macroscale HPM, the conductivity of nanoscale HPM is affected by its relatively larger surface area and quantum effects\cite{bharagava2018recent}. There are many experiments on this issue, such as the electrical conductivity measurement of the graphene hydrogels for supercapacitors\cite{Banda2017,RamosFerrer2017} and nanoporous reduced graphene oxide templated by hydrophobic CaCO3 spheres\cite{Gu2014}. However, numerical calculations of electrical conductivity for nanoscale HPM are rare. Many studies have focused on individual hydrophobic nanopores, such as the conductance calculation for the carbon nanotubes\cite{Secchi2016,Secchi2016b,Uematsu2018} (CNTs), boron nitride nanotubes (BNNTs), and other hydrophobic nanopores with atomic layer deposition\cite{Balme2015} (ALD). 
	
	The effective medium approximation\cite{doyen1988permeability,jurgawczynski2007predicting} (EMA) can help us obtain the conductivity of nanoporous media based on the individual nanopores' calculated conductance. The structure of porous media can be complicated; pore size distribution\cite{Meng2019} and pore connectivity\cite{Li2015} affect the porous media's conductivity.
	
	Therefore, we establish the hydrophobic nanopores with different surface charge densities. A modified Poisson-Boltzmann (MPB) model\cite{Huang2007,Huang2008} is introduced to calculate the ionic charge density profiles that consistent with our equilibrium molecular dynamics (MD) simulation results. The MPB equation and the Stokes equation are combined to calculate the electric current, where the zeta potential\cite{Bocquet2010} is amplified by the slippage that occurs on the hydrophobic surface. Then we calculate the conductance of a set of nanopores with different size in the nanoscale HPM. EMA is involved in calculating the effective conductance and conductivity of the nanoscale HPM\cite{doyen1988permeability}. This research aims to propose a solution to predict the electrical conductivity of a nanoscale HPM with a heterogeneous pore structure.

    \section{Methods}
    \subsection{MD Simulation}
\begin{figure}[h]
    	\centering
    	\includegraphics[width=0.4\textwidth]{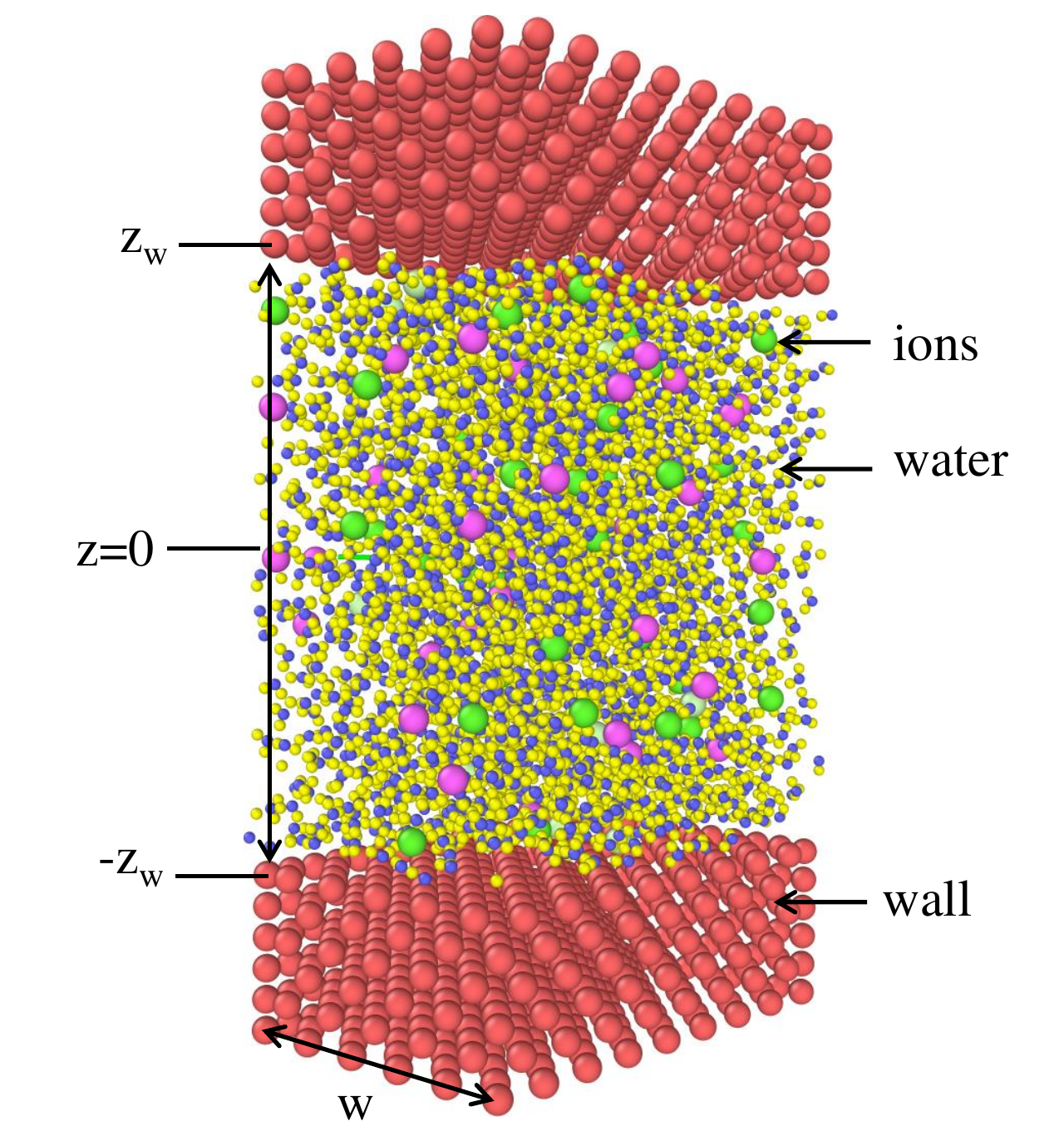}
    	\caption{AES is confined between the walls and is used for the MD simulations and the MPB model. The position of the two hydrophobic surfaces of the walls are ${{z}_{w}}$ and $-{{z}_{w}}$, respectively. $z=0$ is the center position in the vertical direction. The width of the nanopore is $w$.}
    	\label{fig:6}
    \end{figure}

    For a confined AES system (Figure \ref{fig:6}), the gap (50 Å) between the two hydrophobic walls is determined by a preliminary simulation, where the pressure on the wall surface is 10 atm. Each wall is a face-centered cubic system of $9\times 6\times 3$ unit cells, the lattice constant\cite{Fu2019b} is 5.356 Å. We choose a specific set of Lennard-Jones (LJ) parameters\cite{Huang2008,Werder2003} (${{\sigma }_{ss}}=3.374\text{ }$Å, ${{\varepsilon }_{ss}}=0.164\text{ }{\text{kcal}}/{\text{mol}}\;$) for wall atoms to create a hydrophobic surface. For a wall surface with $N$  atoms, a surface area of $S$, and a surface charge density of $\sum $, the charge per atom is $q={\sum S}/{N}\;$. 
    
    The SPC/E (extended simple point charge)\cite{Berendsen1987} water model is used to simulate AES. The two bonds and the angle of the water model is constrained via the SHAKE algorithm\cite{Ryckaert1977}. For a solution with a monovalent salt concentration of 1 mol/L between the walls with no surface charge, the number of water molecules and cations (or anions) are 2080 and 40, respectively. If the wall surface is charged, a certain amount of counterions (${{N}_{c}}=2\times {\sum S}/{{{q}_{c}}}\;$) needs to be added.
    
    The LJ parameters for the ions are listed in Table 5\cite{Joung2008} from Joung and Cheatham (2008). The Lorentz-Berthelot\cite{Lorentz1881,berthelot1898melange} combining rules (${{\sigma }_{ij}}={\left( {{\sigma }_{ii}}+{{\sigma }_{jj}} \right)}/{2}\;$, ${{\varepsilon }_{ij}}=\sqrt{{{\varepsilon }_{ii}}{{\varepsilon }_{jj}}}$) are used for the LJ potential and the cut-off distance is 10 Å. We choose the P3M (particle-particle-particle-mesh) method\cite{hockney1988computer} to calculate the long-range Coulombic forces. A Nose-Hoover thermostat\cite{Posch1986} is used to control the temperature at 298K. The MD simulations are conducted using the LAMMPS\cite{Plimpton1995}. 
    
    \subsection{Modified Poisson-Boltzmann Model}
    If the local cation and anion concentrations in an AES follow a Boltzmann distribution\cite{butt2013physics}, ${{c}^{\pm }}={{c}_{0}}\cdot \exp \left[ \frac{\mp eV\left( z \right)}{{{k}_{B}}T} \right]$, the local charge density, ${{\rho }_{e}}\left( z \right)$, in an AES can be obtained by ${{\rho }_{e}}\left( z \right)=e\left( {{c}^{+}}-{{c}^{-}} \right)$; then, the electric potential distribution, $V\left( z \right)$, can be described by the Poisson-Boltzmann equation, $\frac{{{d}^{2}}V\left( z \right)}{d{{z}^{2}}}=-\frac{e\left( {{c}^{+}}-{{c}^{-}} \right)}{{{\varepsilon }_{0}}{{\varepsilon }_{r}}\left( z \right)}$.
    
\begin{figure}[h]
    	\centering
    	\includegraphics[width=0.4\textwidth]{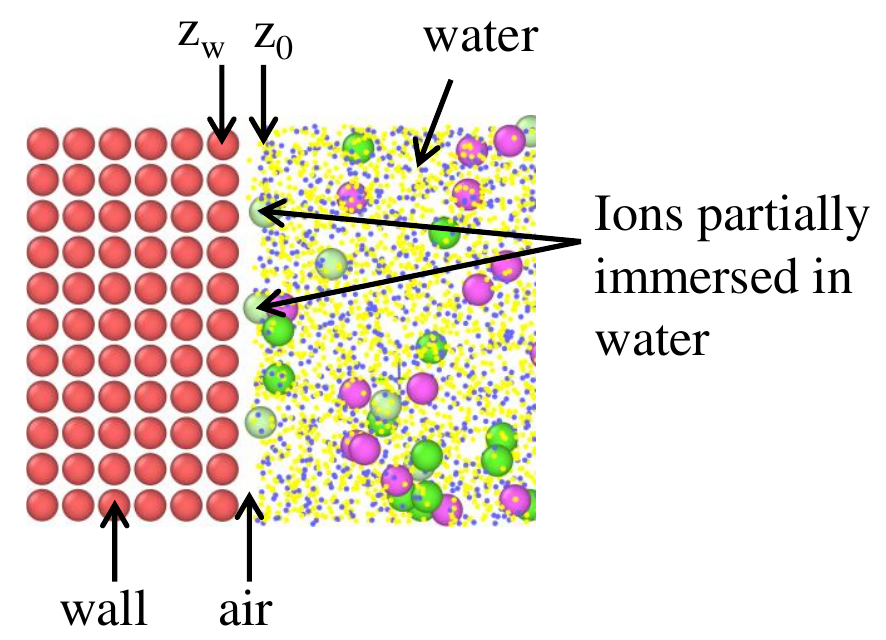}
    	\caption{Part of the ion is in the air layer, and the rest is immersed in water. The hydrophobic solvation energy is related to the cavity volume created by water.}
    	\label{fig:7}
    \end{figure}

    A slight displacement of the bound charges appears when a dielectric material is set in an external electric field, the electric displacement field is given by\cite{Jackson2013}$\mathbf{D}\equiv {{\varepsilon }_{0}}\mathbf{E}+\mathbf{P}$. Combining the Boltzmann distribution, we obtain $\frac{d}{dz}\left[ -{{\varepsilon }_{0}}\frac{d}{dz}V\left( z \right)+P\left( z \right) \right]={{\rho }_{e}}\left( z \right)$, where ${{\rho }_{e}}\left( z \right)=e\left( {{c}^{+}}-{{c}^{-}} \right)$, ${{c}^{\pm }}={{c}_{0}}\cdot \exp \left[ \frac{\mp eV\left( z \right)}{{{k}_{B}}T} \right]$, and $P\left( z \right)={{\varepsilon }_{0}}{{\varepsilon }_{r}}\left( z \right)E-{{\varepsilon }_{0}}E$. Therefore,	
    \begin{equation} \label{eq:1}
    	-{{\varepsilon }_{0}}\frac{d}{dz}\left[ {{\varepsilon }_{r}}\left( z \right)\frac{d}{dz}V\left( z \right) \right]={{\rho }_{e}}\left( z \right)
    \end{equation}
    which is very similar to the form of the Poisson-Boltzmann equation and is called the “step-polarization model”\cite{Huang2008}, and the relative permittivity, ${{\varepsilon }_{r}}\left( z \right)$, is given by\cite{Huang2008,Fumagalli2018},
    \begin{equation} \label{eq:2}
    	{{\varepsilon }_{r}}=\left\{ \begin{matrix}
    		1 & {{\text{z}}_{0}}<\text{z}<{{\text{z}}_{w}}  \\
    		{h}/{\left[ 2{{{h}_{i}}}/{{{\varepsilon }_{i}}+{\left( h-2{{h}_{i}} \right)}/{{{\varepsilon }_{bulk}}}\;}\; \right]}\; & \text{z}\le {{\text{z}}_{0}}  \\
    	\end{matrix} \right.		
    \end{equation}
    where $h=2{{z}_{0}}$, ${{h}_{i}}=7.4$ Å, ${{\varepsilon }_{i}}=2.1$ Å, ${{\varepsilon }_{bulk}}=68$. The thickness of the air layer is equal to ${{z}_{w}}-{{z}_{0}}$. ${{z}_{0}}$ can be determined according to the water density profiles of MD simulations\cite{Huang2008} (Figure \ref{fig:7}).
    
    The appearance of air-water interface complicates the change in the concentration of the aqueous electrolyte solution near the hydrophobic wall\cite{Bostrom2005}. Adding an external potential, $U_{\text{ext}}^{\text{ }\!\!\pm\!\!\text{ }}\left( z \right)$, to the Boltzmann equation, ${{c}^{\pm }}={{c}_{0}}\cdot \exp \left[ \frac{\mp eV\left( z \right)-U_{\text{ext}}^{\text{ }\!\!\pm\!\!\text{ }}\left( z \right)}{{{k}_{B}}T} \right]$, can help us describe the concentration profile more accurately. The external potential is represented by three potential terms\cite{Huang2008},
    \begin{equation} \label{eq:3}
    	U_{\text{ext}}^{\text{ }\!\!\pm\!\!\text{ }}\left( z \right)\text{=}U_{\text{image}}^{\pm }\left( z \right)\text{+}U_{\text{wall}}^{\pm }\left( z \right)\text{+}U_{\text{solvation}}^{\pm }\left( z \right)
    \end{equation}
    
    According to Wagner (1924), the ions near the air-water interface are affected by the “image force”\cite{wagner1924surface,Onsager1934}, which will cause changes in the concentration of AES near the interface\cite{Bostrom2001}. The potential of this force has the form,
    \begin{equation} \label{eq:4}
    	U_{\text{image}}^{\pm }\left( z \right)=\left( \frac{{{\varepsilon }_{r}}-1}{{{\varepsilon }_{r}}+1} \right)\frac{{{e}^{2}}\exp \left[ -2\kappa \left( z-{{z}_{0}} \right) \right]}{16\pi {{\varepsilon }_{0}}{{\varepsilon }_{r}}\left( z-{{z}_{0}} \right)}
    \end{equation}
    where ${{\kappa }^{-1}}=\sqrt{{{{\varepsilon }_{0}}{{\varepsilon }_{r}}{{k}_{B}}T}/{\left( 2{{e}^{2}}{{c}_{0}} \right)}\;}$ is the Debye length.
    
    The wall is assumed to be fully homogeneous in both x and y directions. The total potential energy of interaction between an ion and the wall is obtainable by integration over x and y\cite{Huang2008,zhao2012simple}. The distance between the two along the z direction is $z-{{z}_{w}}$.
    \begin{equation} \label{eq:5}
    	\begin{split}
    		U_{\text{wall}}^{\pm }\left( z \right) &=\frac{2}{3}\pi \rho _{s}^{*}{{\varepsilon }_{s\pm }}\times \\
    		& \left[ \frac{2}{15}{{z}^{*}}^{-9}-{{z}^{*}}^{-3}+{{z}^{*}}\left( \frac{36}{5}r_{c}^{*-10}-9r_{c}^{*-4} \right) \right.\left. +2{{z}^{*3}}\left( -r_{c}^{*-12}+r_{c}^{*-6} \right)-\frac{16}{3}r_{c}^{*-9}+8r_{c}^{*-3} \right]
    	\end{split}	
    \end{equation}
    where $\rho _{s}^{*}={{\rho }_{s}}\sigma _{s\pm }^{3}$, ${{z}^{*}}=\left( z-{{z}_{w}} \right)/{{\sigma }_{s\pm }}$, and $r_{c}^{*}={{{r}_{c}}}/{{{\sigma }_{s\pm }}}$. This potential is equal to zero when $z-{{z}_{w}}\ge {{r}_{c}}$.
    
    Before an ion enters the water, a cavity must be created so that the water has space to hold the ion. According to Israelachvili (2011), this process results in a “many-body interaction”\cite{israelachvili2011intermolecular} that occurs primarily between water molecules in the AES. The relationship between the cavity volume changes and the hydrophobic solvation energy\cite{Huang2008} is given by,
    \begin{equation} \label{eq:6}
    	U_{\text{solvation}}^{\pm }\left( z \right)={{C}_{0}}\left[ v_{\text{cavity}}^{\pm }\left( z \right)-v_{\text{ion}}^{\pm } \right]
    \end{equation}
    where the cavity volume (Figure \ref{fig:7}), that is, the volume excluded by the ion in the water, is $v_{\text{cavity}}^{\pm }\left( z \right)={\pi \left( 3{{r}_{\pm }}z_{\pm }^{2}-z_{\pm }^{3} \right)}/{3}\;$, ${{z}_{\pm }}=z-{{z}_{0}}+{{r}_{\pm }}$, and $0\le {{z}_{\pm }}\le 2{{r}_{\pm }}$. The ion volume is $v_{\text{ion}}^{\pm }\left( z \right)={4\pi r_{\pm }^{3}}/{3}\;$. For the NaI AES, the radii of sodium ions and iodide ions are set to 1.54 Å and 3.2 Å, respectively. The proportionality factor is ${{C}_{0}}=2.8\times {{10}^{8}}{\text{J}}/{{{\text{m}}^{\text{3}}}}\;$. We use the FEniCS\cite{Logg2010,logg2012automated} to solve the MPB model (Eq \ref{eq:1}) for the ionic charge density profiles, ${{\rho }_{e}}\left( z \right)$.
    
    \subsection{Current and Conductance}
    When an electric force is applied as a body force in the x-direction parallel to the wall surface, the equation for the Stokes flow is given by\cite{Bocquet2010,leal2007advanced},
    \begin{equation} \label{eq:7}
    	\eta \frac{{{\partial }^{2}}v}{\partial {{z}^{2}}}+{{\rho }_{e}}\left( z \right)E=0
    \end{equation}
    
    Combining the Poisson equation and the two boundary conditions, ${v}'\left( 0 \right)=0$ and $b{v}'\left( {{z}_{0}} \right)=v\left( {{z}_{0}} \right)$ ($b$ is the slip length), we can obtain the fluid velocity profile,
    \begin{equation} \label{eq:8}
    	v\left( z \right)=-\frac{\varepsilon }{\eta }E\left[ -V\left( z \right)+\zeta  \right]
    \end{equation}
    where the zeta potential is $\zeta ={{V}_{0}}\left( 1+b{{\kappa }_{\text{eff}}} \right)$, the surface screening parameter is ${{\kappa }_{\text{eff}}}={-{V}'\left( 0 \right)}/{{{V}_{0}}}\;$ (${{V}_{0}}$ is the electrostatic potential for the hydrophobic surface). In this study, we used a formula related to slip length and surface charge density to calculate the zeta potential, \\
    $\zeta =-\frac{1}{{{\varepsilon }_{0}}{{\varepsilon }_{w}}}\int_{-{{z}_{0}}}^{0}{dz\left( z+{{z}_{0}}+b \right){{\rho }_{e}}\left( z \right)}$, where the slip length can be calculated using \\
    $b={{b}_{_{0}}}\left[ 1+\left( {1}/{\alpha }\; \right){{\left( {\sum {{\sigma }^{2}}}/{e}\; \right)}^{2}}\left( {{{l}_{B}}}/{\sigma }\; \right)\left( {{{b}_{0}}}/{\sigma }\; \right) \right]$, $b\left( \sum =0 \right)={{b}_{_{0}}}$, the numerical prefactor is $\alpha \sim 1.7$, $\sigma ={{\sigma }_{ss}}=3.374$ Å, ${{l}_{B}}={{{e}^{2}}}/{\left( 4\pi {{\varepsilon }_{0}}{{\varepsilon }_{surf}}{{k}_{B}}T \right)}\;$, the relative permittivity at the liquid – solid surface\cite{Huang2008} is ${{\varepsilon }_{surf}}\approx 1$.
    
    The average velocity of ions can be calculated from the fluid velocity and the drift velocity, ${{u}_{\pm }}\left( z \right)=v\left( z \right)\pm e{{\mu }_{\pm }}E$, where the ion mobility is ${{\mu }_{\pm }}={e}/{6\pi \eta {{r}_{\pm }}}$. Then the current can be calculated with the following expression\cite{Bocquet2010},
    \begin{equation} \label{eq:9}
    	I=we\int_{-{{z}_{w}}}^{{{z}_{w}}}{dz\left[ {{c}^{+}}\left( z \right){{u}_{+}}\left( z \right)-{{c}^{-}}\left( z \right){{u}_{-}}\left( z \right) \right]}
    \end{equation}
    where the width of the nanopore is $w$. The conductance of the nanopore can be represented as $G={I}/{V}\;$, where the voltage is $V$.
    
    \subsection{EMA and Conductivity}
    The electrical conductivity of the AES filling a throat can be calculated as ${{\sigma }_{f}}\text{=}{GL}/{A}\;$, where $L$ is the length of the throat, $A$ is the cross-sectional area of the throat ($A=\pi {{r}^{2}}={{w}^{2}}$). The conductivity, ${{\sigma }^{*}}$, of the nanoscale HPM can be obtained from the effective medium approximation (EMA)\cite{doyen1988permeability},
    \begin{equation} \label{eq:10}
    	\sum\limits_{r}{{{n}_{t}}\left( r \right)\frac{{{G}^{*}}\left( {{r}^{*}} \right)-G\left( r \right)}{G\left( r \right)+\left( {{{z}_{t}}}/{2}\;-1 \right){{G}^{*}}\left( {{r}^{*}} \right)}}=0
    \end{equation}
    where $r$ and ${{r}^{*}}$ are the throat radius and characteristic throat radius, respectively. The latter, ${{r}^{*}}$, is a calculation result of EMA. ${{n}_{t}}\left( r \right)$ is the throat size distribution, ${{G}^{*}}$ is the effective conductance of any throat in an hypothetical homogeneous porous media with conductivity ${{\sigma }^{*}}$ equal to that of the actual HPM with heterogeneous pore structure. The coordination number, ${{z}_{t}}$, is equal to the number of branches (throats) meeting at each node (pore chamber). 
    We use Newton–Raphson method\cite{jurgawczynski2007predicting} to find successive approximations to ${{G}^{*}}$ and ${{r}^{*}}$. Then the electrical conductivity, ${{\sigma }^{*}}$, is given by\cite{doyen1988permeability},
    \begin{equation} \label{eq:11}
    	{{\sigma }^{*}}\approx \frac{1}{\tau }{{\sigma }_{f}}\phi \frac{{{r}^{*2}}}{\left\langle r_{p}^{2} \right\rangle }
    \end{equation}
    where $\tau$ and $\phi$ are the tortuosity and porosity of the actual nanoscale HPM, respectively. The average of $r_{p}^{2}$ over the pore size distribution of the HPM is $\left\langle r_{p}^{2} \right\rangle $.
    
	\section{Results and Discussion}
	\subsection{Ionic Charge Density Profile}
\begin{figure}[h]
 	  \centering
	  \begin{subfigure}{0.4\textwidth}
		\includegraphics[width=0.9\textwidth, height=1.8in]{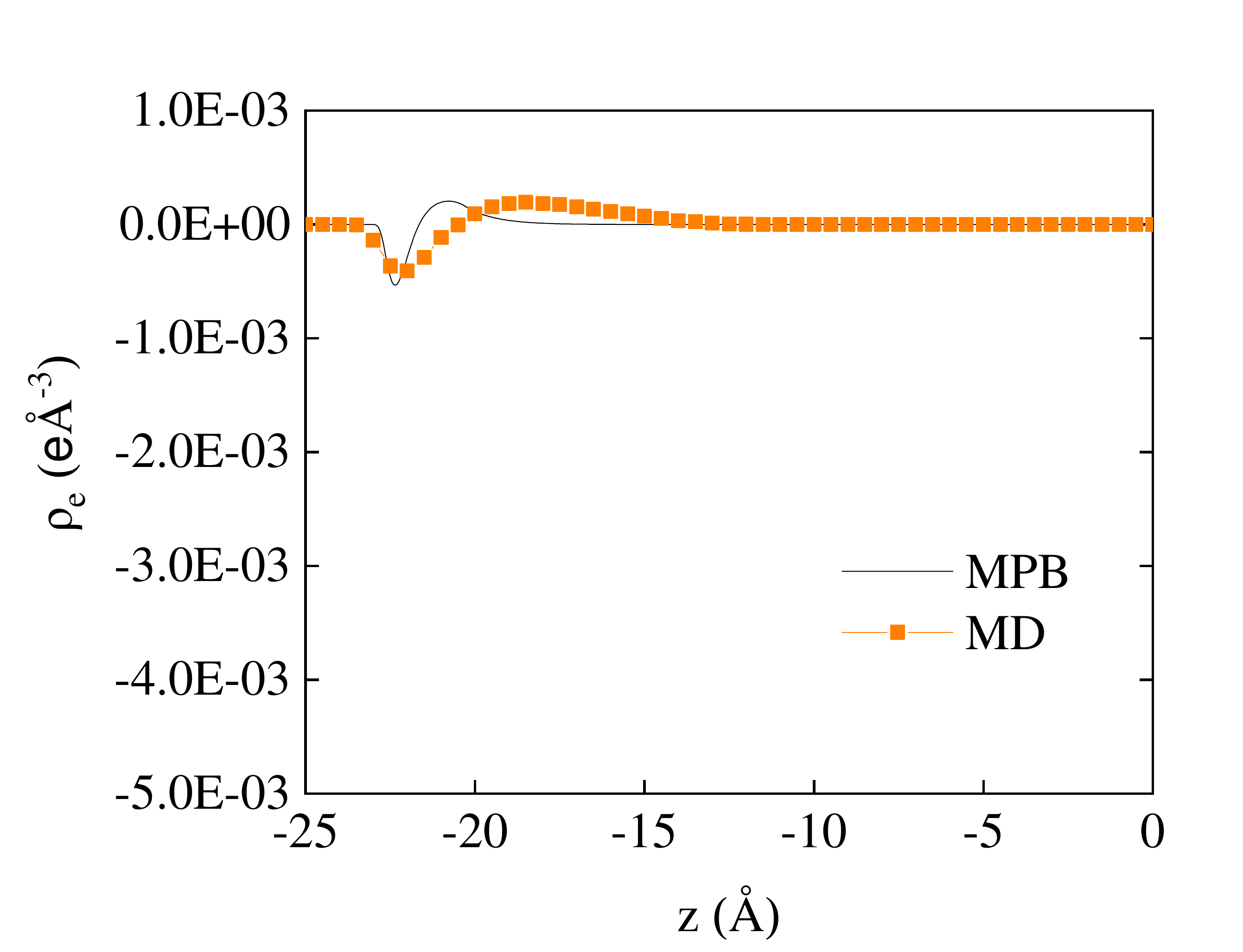}
		\caption{\label{fig:1a}}
	  \end{subfigure}
	  \begin{subfigure}{0.4\textwidth}
		\includegraphics[width=0.9\textwidth, height=1.8in]{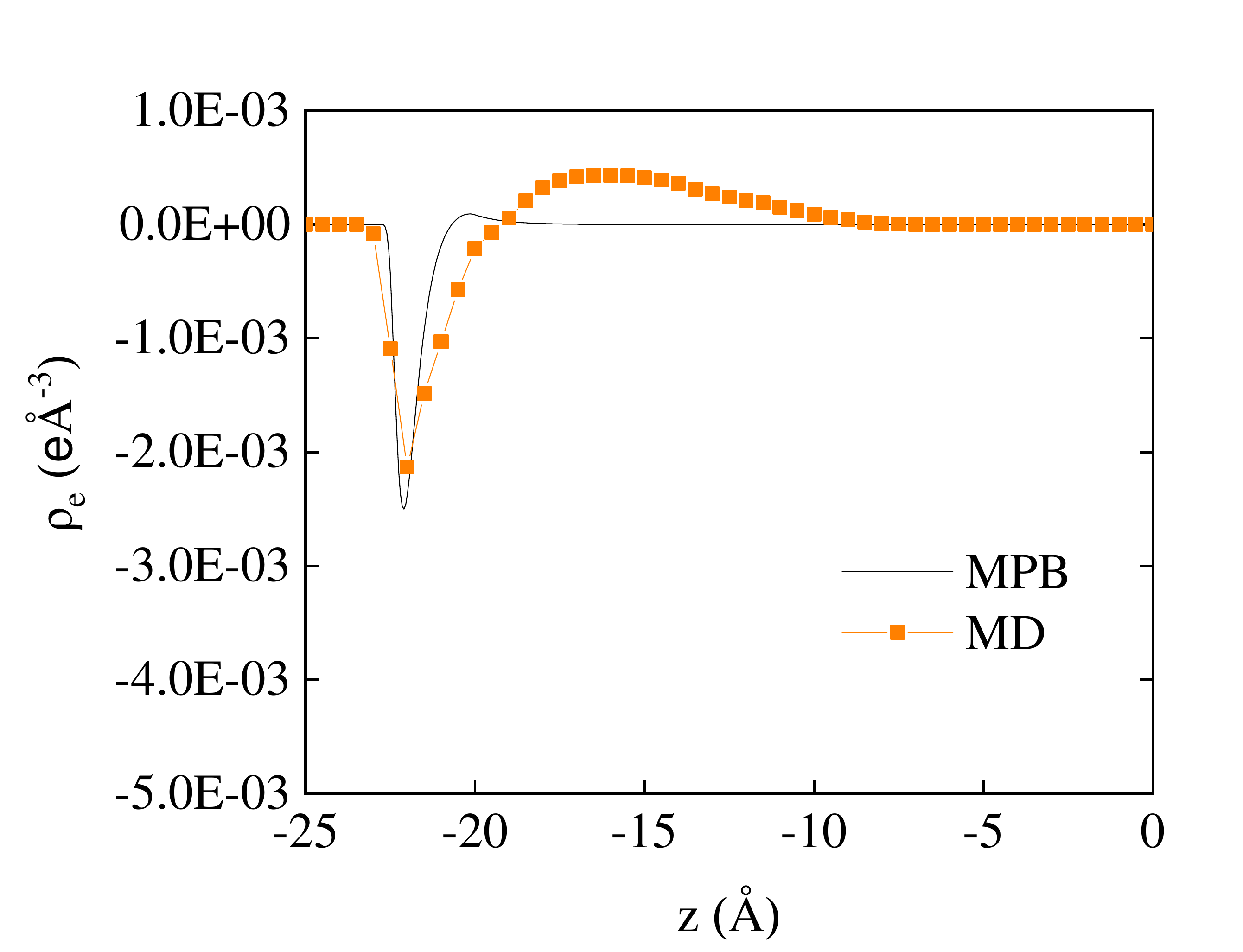}
		\caption{\label{fig:1b}}
	  \end{subfigure}
	  \begin{subfigure}{0.4\textwidth}
		\includegraphics[width=0.9\textwidth, height=1.8in]{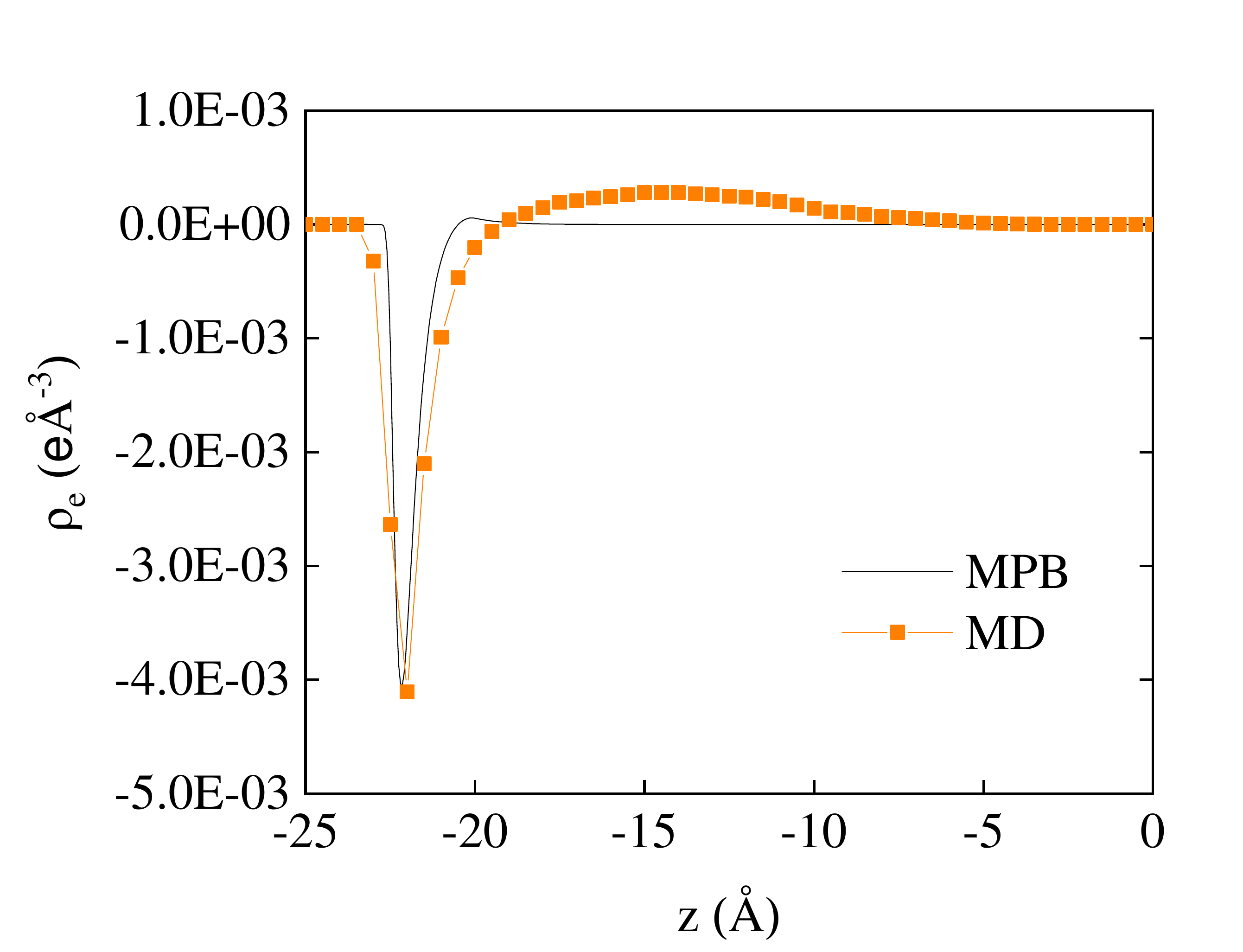}
		\caption{\label{fig:1c}}
	  \end{subfigure}
	  \caption{The ionic charge density profiles produced by the MPB model are compared with the MD results. The molarity of the NaI aqueous electrolyte solution (AES) is 1 M. The width of the nanopores is 5 nm. The surface charge density of the hydrophobic nanopores is 0 (\subref{fig:1a}), +0.031 C/m$^2$ (\subref{fig:1b}), and +0.062 C/m$^2$ (\subref{fig:1c}).}
	  \label{fig:1}
    \end{figure}

	The ionic charge density profiles generated by the MPB model and the MD simulations show consistency (Figure \ref{fig:1}). After the relative permittivity expression of confined water\cite{Fumagalli2018} (Eq \ref{eq:2}) and the external potentials\cite{Huang2008} (Eq \ref{eq:3}) are introduced, the ionic charge density profile can be described in more detail by the MPB model (Eq \ref{eq:1}). 
	
\begin{figure}[H]
		\centering
		\includegraphics[width=0.4\textwidth]{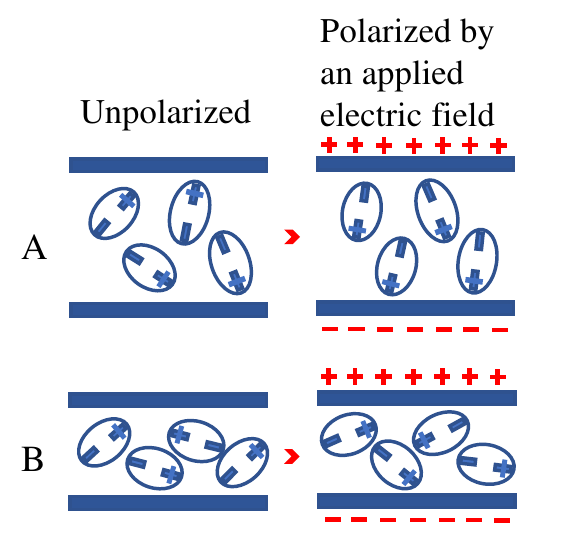}
		\caption{Compared with nanopore A, nanopore B has extreme confinement. When an electric field is applied, water dipoles in nanopore B feel more challenging to reorient. The polarizability of confined water decreases as the dipole rotational freedom decreases.}
		\label{fig:2}
	\end{figure}

	Experimental evidence\cite{Fumagalli2018} shows that the relative permittivity of water confined in the nanopores is abnormally low. There are many nanopores of different sizes in a nanoscale HPM. The water thickness in a smaller hydrophobic nanopore is also thinner. As a result, the electric polarizability of confined water decreases (Figure \ref{fig:2}), and its relative permittivity decreases.	
	
	Compared with a sodium ion, an iodide ion has a larger radius, which means that water needs to create a larger cavity to accommodate an iodide ion and consume more energy\cite{israelachvili2011intermolecular}. The hydrophobic solvation energy\cite{Huang2008,Bostrom2005}, $U_{\text{solvation}}^{\pm}$, helps quantify the energy change when the ions move near the air-water interface. Together with the image potential, $U_{\text{image}}^{\pm }$, and the ion-wall potential, $U_{\text{wall}}^{\pm }$, the MPB model included these external potentials can accurately describe the presence of more iodide ions near the wall, especially when the surface charge density of the hydrophobic nanopore increases (Figure 3\subref{fig:1c}).

	\subsection{Conductance of Individual Hydrophobic Nanopores}
\begin{figure}[H]
	  \centering
	  \begin{subfigure}{0.4\textwidth}
		\includegraphics[width=0.9\textwidth, height=1.8in]{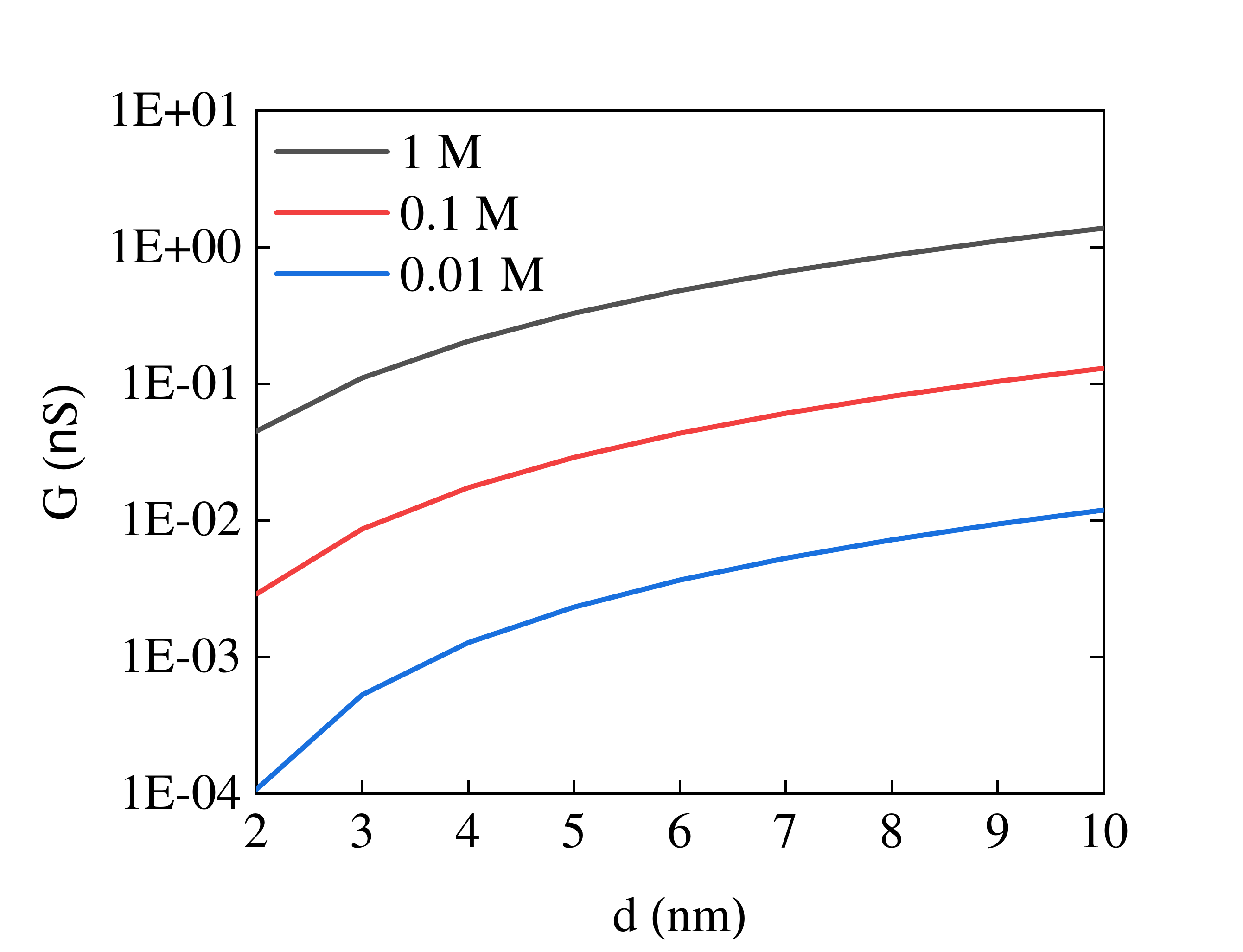}
		\caption{\label{fig:3a}}
  	  \end{subfigure}
	  \begin{subfigure}{0.4\textwidth}
		\includegraphics[width=0.9\textwidth, height=1.8in]{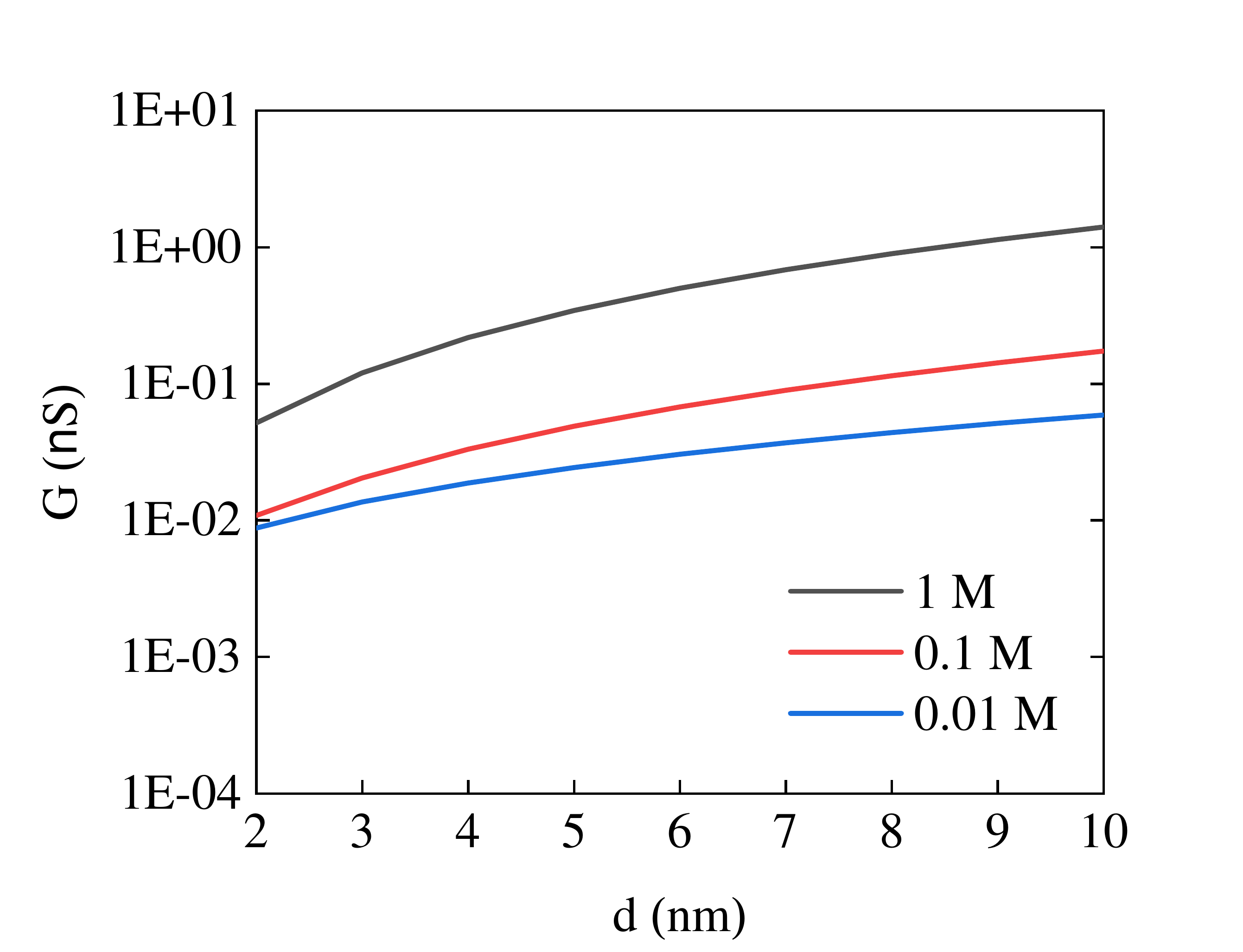}
		\caption{\label{fig:3b}}
	  \end{subfigure}
	  \begin{subfigure}{0.4\textwidth}
		\includegraphics[width=0.9\textwidth, height=1.8in]{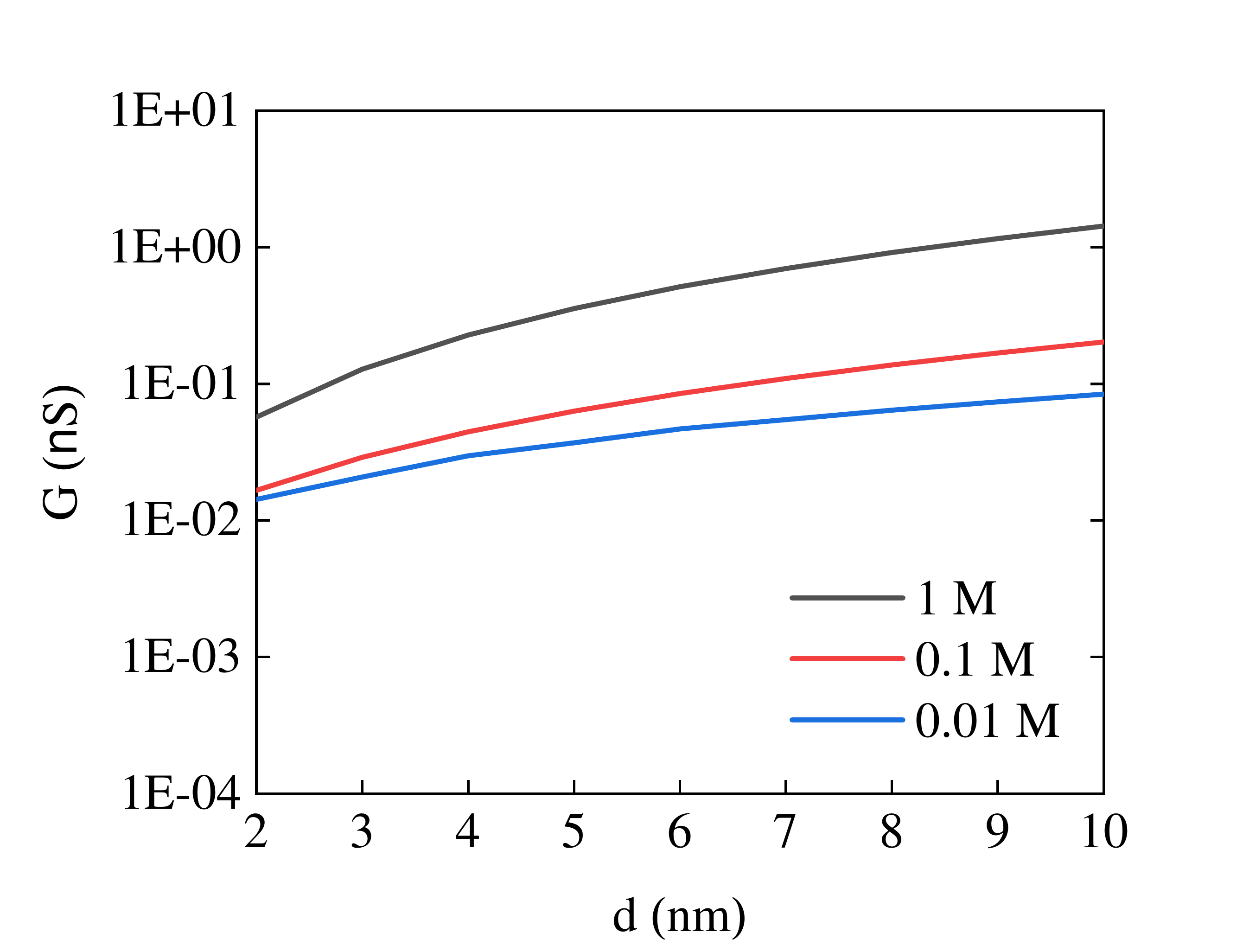}
		\caption{\label{fig:3c}}
	  \end{subfigure}
	  \caption{Conductance versus pore size of hydrophobic nanopores. The molarity of the NaI AES is changed from 0.01 M to 1 M. The pore size is changed from 2 nm to 10 nm. The surface charge density of the hydrophobic nanopores is 0 (\subref{fig:3a}), +0.031 C/m$^2$ (\subref{fig:3b}), and +0.062 C/m$^2$ (\subref{fig:3c}). There is no Debye layer overlap in all cases.}
	  \label{fig:3}
    \end{figure}

	The calculation results based on the combination of the MPB model and the Stokes equation (Figure \ref{fig:3}). The pore size (width and height) of each hydrophobic nanopore is w, the length is 900 nm, the voltage applied to the nanopores is 1 V. When the pore size increases, the conductance increases. When the molarity of the aqueous electrolyte solution (AES) decreases, the conductance decreases, and the magnitude of the decrease also gradually slows down (Figure 5\subref{fig:3b} and 5\subref{fig:3c}), which is related to the conductance saturation\cite{Bocquet2010,PhysRevLett.93.035901}. 
	
	According to Bocquet (2010), the race between surface conductance and bulk conductance can be evaluated by the “Dukhin length” ${{l}_{Du}}={{\left| \sum  \right|}/{e}\;}/{{{c}_{0}}}$. For the case where the surface charge density is $\sum =+0.062\text{ C/}{{\text{m}}^{\text{2}}}$ and the molarity of the AES is ${{c}_{0}}=0.1\text{ M}$, ${{l}_{Du}}\approx 6.4\text{ nm}$. Therefore, in this case, the surface contribution to the nanopore conductance is dominant\cite{Bocquet2010} when the pore size is smaller than 6.4 nm.
	
	When the hydrophobic nanopore has a neutral surface (Figure 5\subref{fig:3a}), the curve with a 1 M molarity is like the corresponding curve in Figure 5\subref{fig:3b} and 5\subref{fig:3c}. However, when the molarity drops to 0.01M, the conductance decreases significantly due to the lack of the contribution of surface conductance.
	
	\subsection{Electrical Conductivity of Nanoscale HPM}
\begin{figure}[h]
		\centering
		\begin{subfigure}{0.4\textwidth}
		\includegraphics[width=0.9\textwidth, height=1.8in]{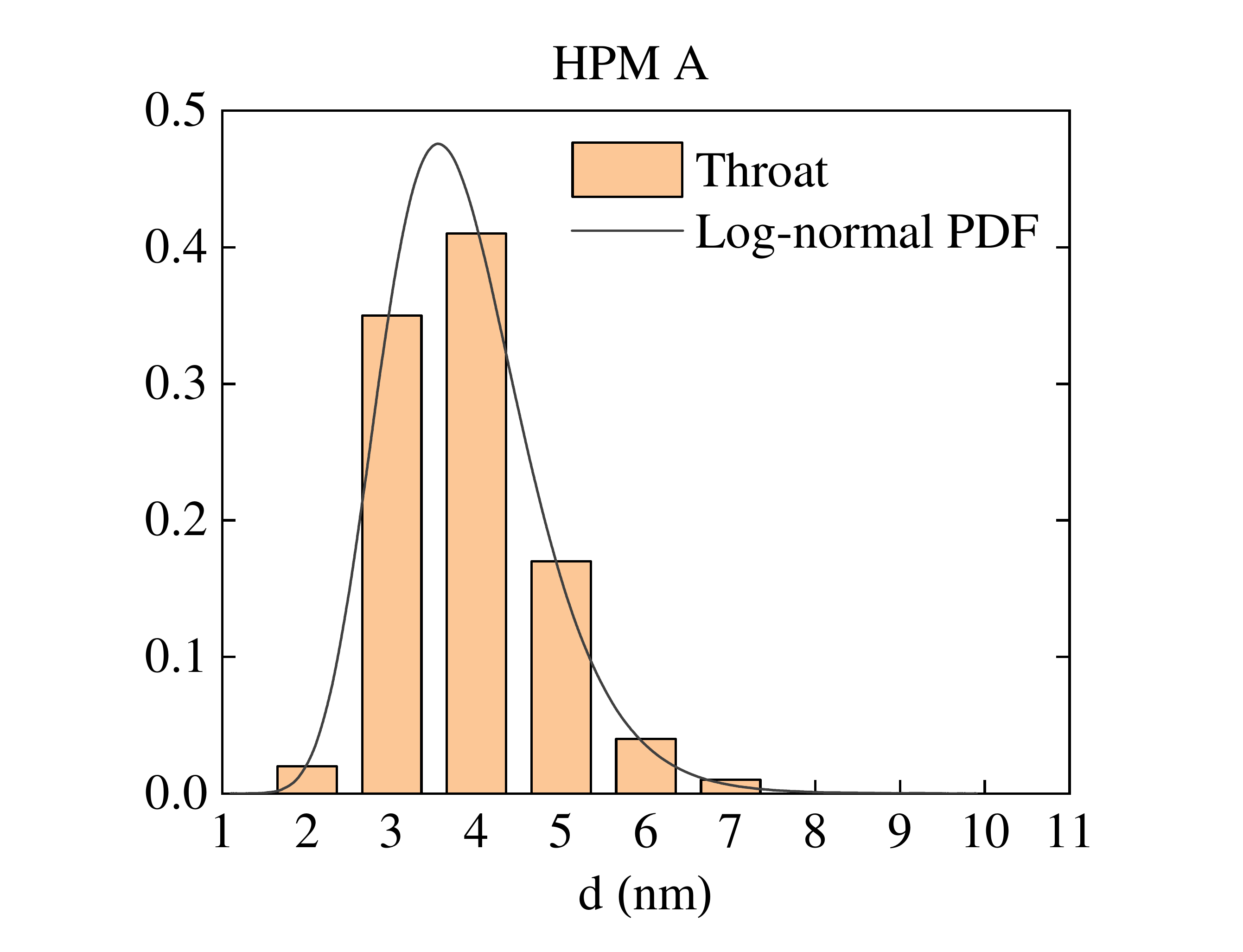}
		\caption{\label{fig:4a}}
		\end{subfigure}
		\begin{subfigure}{0.4\textwidth}
		\includegraphics[width=0.9\textwidth, height=1.8in]{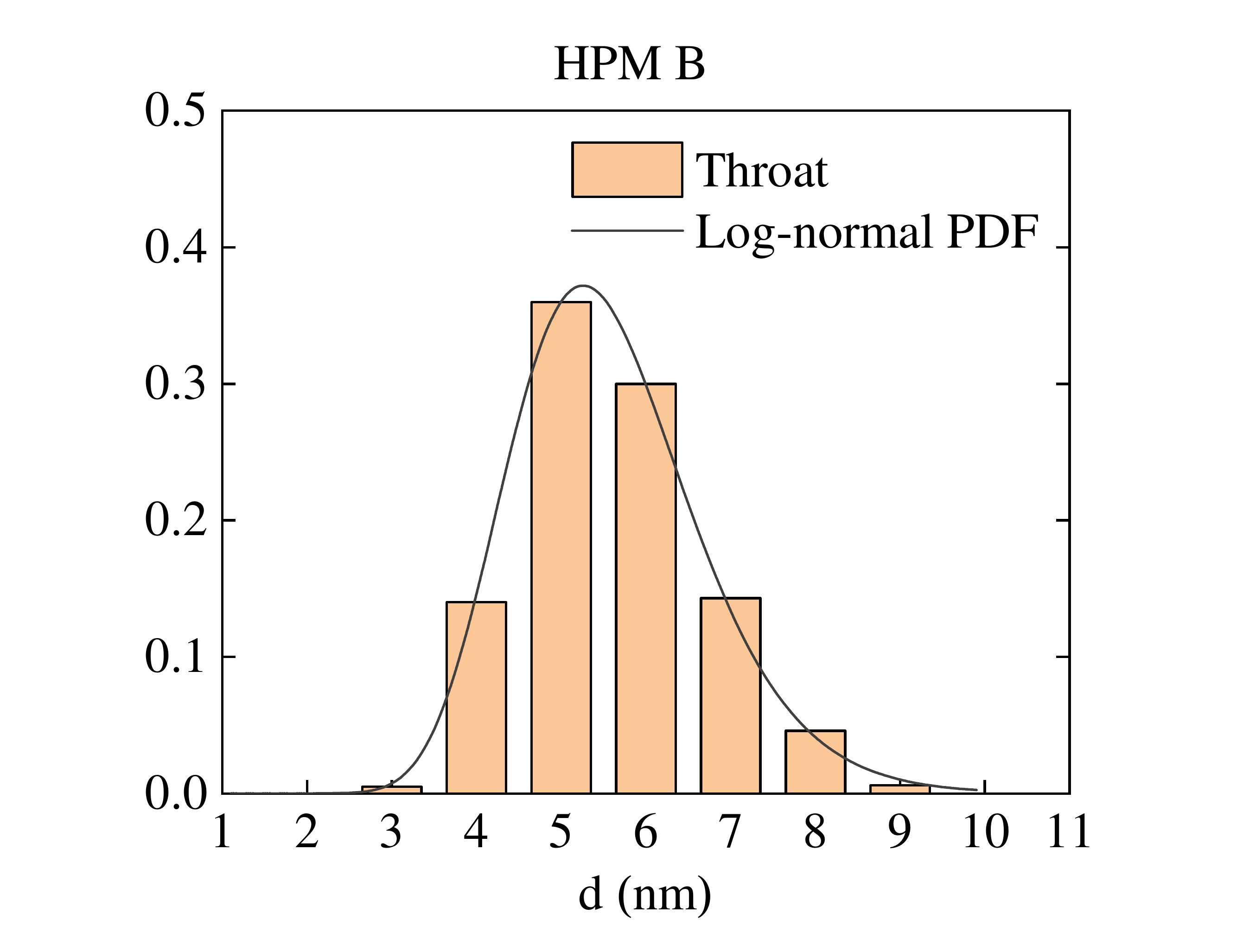}
		\caption{\label{fig:4b}}
		\end{subfigure}
		\begin{subfigure}{0.4\textwidth}
		\includegraphics[width=0.9\textwidth, height=1.8in]{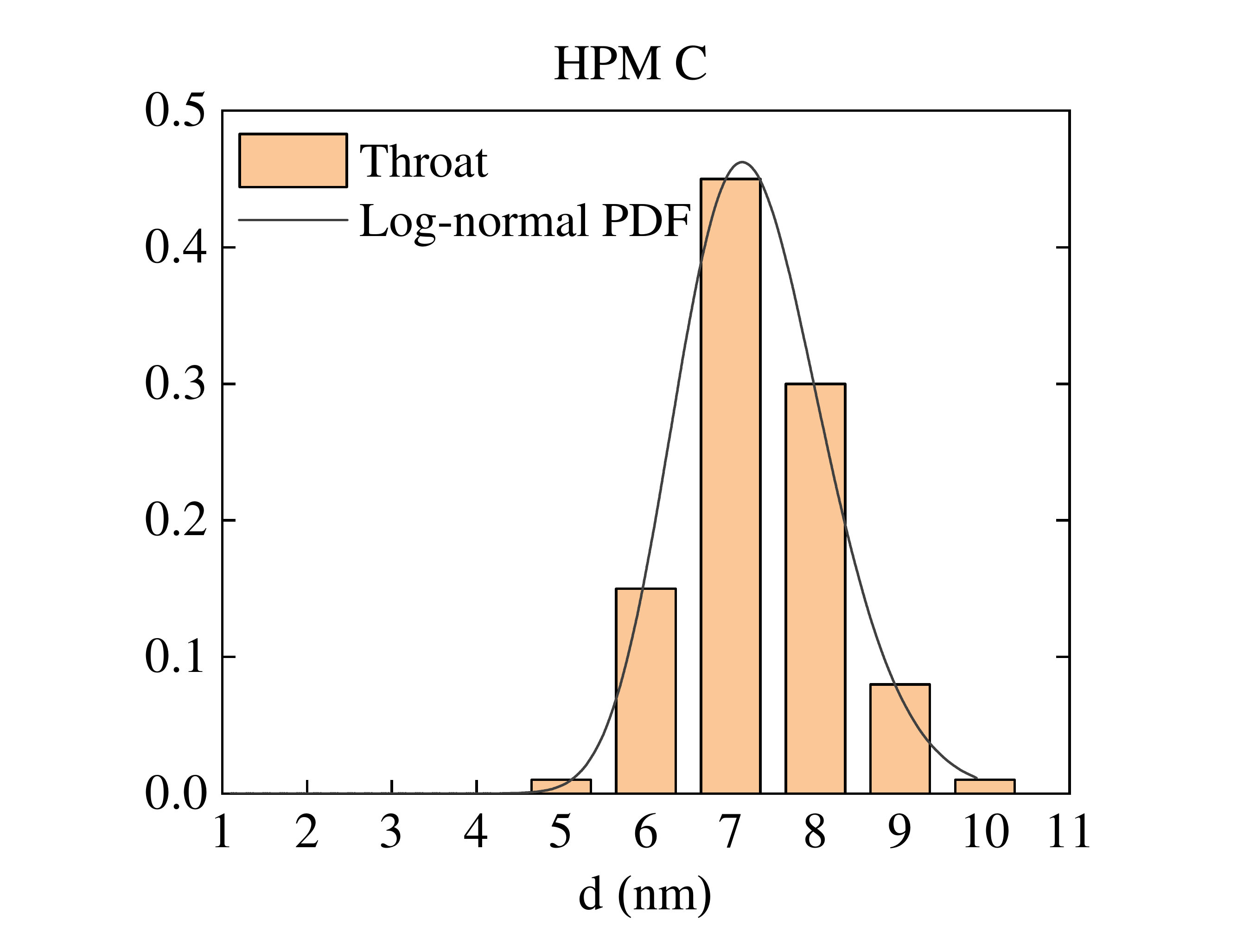}
		\caption{\label{fig:4c}}
		\end{subfigure}
		\caption{Compare the different throat size distributions of three types of nanoscale HPM (A, B, and C). The log-normal PDFs’ median is equal to 3.74 nm, 5.47 nm, and 7.24 nm, respectively. The parameters $\left( \mu ,\text{ }\sigma  \right)$ of these PDFs are (1.32, 0.23), (1.7, 0.2), (1.98, 0.12), respectively.}
		\label{fig:4}
    \end{figure}

	In order to calculate and analyze the electrical conductivity of nanoscale HPM affected by the heterogeneous pore structure without the support of experimental data, we set three types of log-normal probability density function (PDF), $1/\left( d\sigma \sqrt{2\pi } \right)\exp \left[ -{{\left( \ln d-\mu  \right)}^{2}}/\left( 2{{\sigma }^{2}} \right) \right]$, and reference these PDFs to set different throat size distributions (Figure \ref{fig:4}). 
	
\begin{figure}[h]
		\centering
		\includegraphics[width=0.5\textwidth]{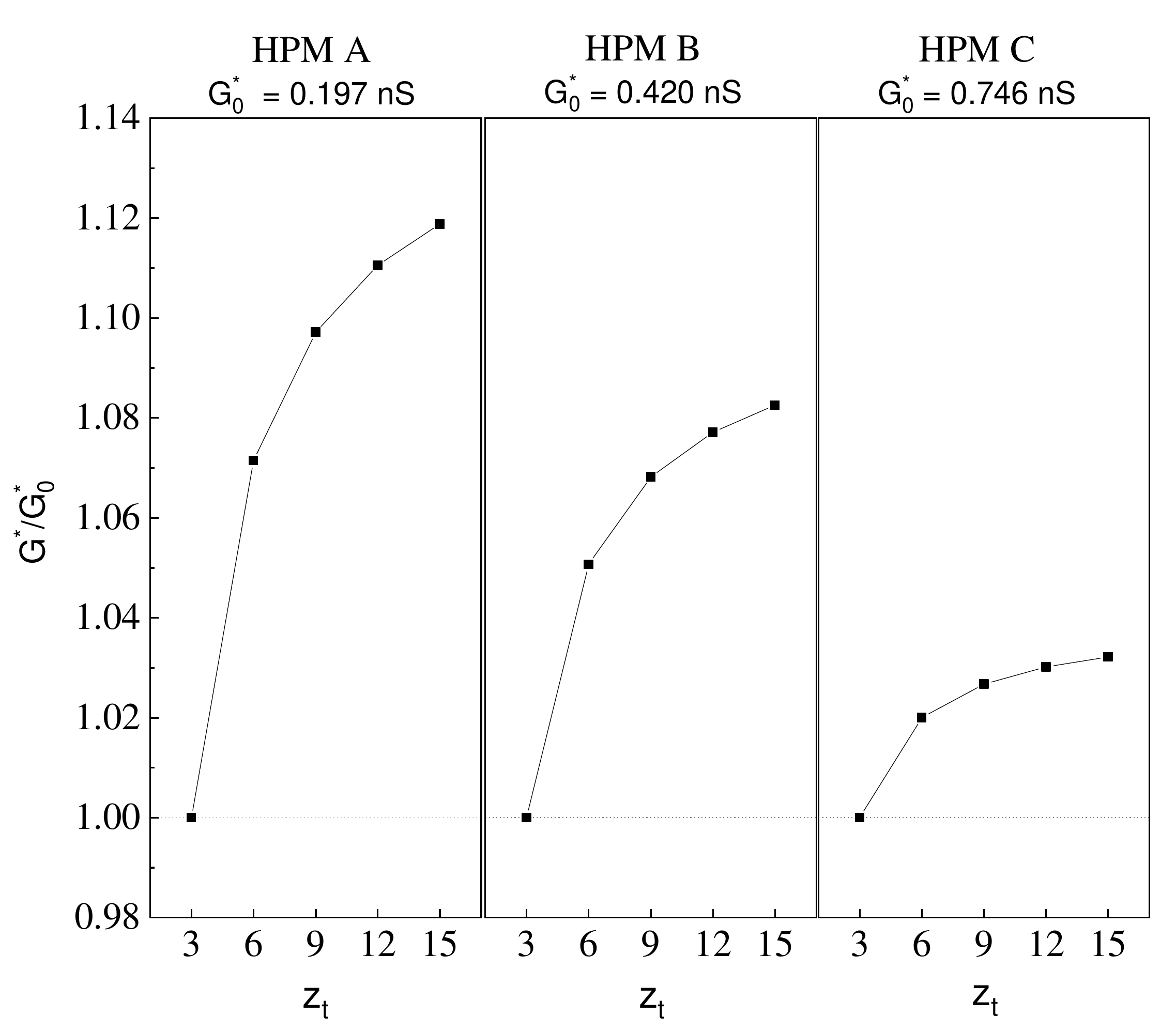}
		\caption{Compare the effective conductance of the three types of nanoscale HPM (A, B, and C) calculated using EMA. The coordination number of the heterogeneous pore structure of the nanoscale HPM is changed. The surface charge density of the nanopores of the HPM is +0.062 C/m$^2$. The molarity of the NaI AES in the HPM is 1 M.}
		\label{fig:5}
	\end{figure}

	The effective conductance, ${{G}^{*}}$, of the three types of nanoscale HPM (A, B, and C) are calculated separately using the EMA (Eq \ref{eq:10}). We change the coordination number,  ${{z}_{t}}$, of the heterogeneous pore structure of the nanoscale HPM and find an impressive result (Figure \ref{fig:5}). That is, for the throat size distribution of HPM with a higher median value (HPM C), the effective conductance of HPM is larger. However, when the coordination number increases, i.e., the pore connectivity of the heterogeneous pore structure becomes better, the increase in the effective conductance of the nanoscale HPM is reduced (HPM A $>$ HPM B $>$ HPM C).
	
	Then we can use Eq \ref{eq:11} to calculate the electrical conductivity of the nanoscale HPM. For actual nanoscale HPM, the porosity and tortuosity should be considered and the throat size distributions (for pore necks), the pore size distributions (for pore chambers). For example, the porosity of one of the nanoscale HPM C is 30, the tortuosity is 3, the coordination number is 6, and the average of $r_{p}^{2}$ over the pore size distribution of the nanoscale HPM, $\left\langle r_{p}^{2} \right\rangle $, is 24.0 nm. Then according to the effective conductance, ${{G}^{*}}$, of HPM C in Figure \ref{fig:5}, we can get the corresponding characteristic throat radius, ${{r}^{*}}$, of 4.1 nm, and the final calculated electrical conductivity of this nanoscale HPM, ${{\sigma }^{*}}$, should be 0.85 nS/nm.

	\section{Conclusions}
	We developed the MPB model to generate the ionic charge density profiles, the relative permittivity expression of confined water, and the external potentials were involved. The surface charge density of the hydrophobic nanopores is changed. The results of the MPB model are consistent with the results of the MD simulations. Then, we calculated the conductance of individual hydrophobic nanopores with different pore sizes. The conductance increases with the increase of the pore size. The conductance decreases with the decrease of the AES molarity. The conductance results also show the contribution of the hydrophobic nanopores’ surface conductance, which is related to conductance saturation.
	
	Then, we set three types of throat size distributions and the corresponding nanoscale HPM (A, B, and C). We changed the coordination number of these nanoscale HPM and used EMA to calculate the effective conductance separately. The effective conductance increases as the coordination number increases. However, for HPM C with a larger average throat size, the effective conductance increase is reduced. Finally, we calculated the electrical conductivity of the nanoscale HPM based on the calculated effective conductance, porosity, tortuosity, and other pore structure parameters.
	
	The above solution can be used to evaluate the transport properties of rocks, supercapacitors, and other materials with nanoporous structures. For a real nanoscale HPM, if the experimental data of heterogeneous pore structure and the electrical conductivity are available, it can be compared with our theoretical calculations. For the current MPB model, the ion specificity effect can be involved\cite{Huang2008}. Surface conductance can generate dynamical selectivity in ion transport\cite{Poggioli2019}. The flexible improvement space\cite{Bostrom2005} of the MPB model is conducive to handling other special situations.
	
  	\section*{Acknowledgments}
	The authors gratefully acknowledge support from the EPSRC “Frontier Engineering” Centre for Nature Inspired Engineering (EP/K038656/1) and an EPSRC “Frontier Engineering: Progression” Grant (EP/S03305X/1).


\bibliographystyle{IEEEtran}
\bibliography{library.bib}


\end{document}